\documentclass[a4paper,11pt,reqno]{amsart}
\usepackage{color}
\usepackage{amssymb,amsmath}
\usepackage{mathrsfs}

\newtheorem{thm}{Theorem}[section]

\newtheorem{lem}[thm]{Lemma}
\newtheorem{dfn}[thm]{Definition}
\newenvironment{defn}{\begin{dfn} \rm }{\end{dfn}}
\newtheorem{cor}[thm]{Corollary}
\newtheorem{example}[thm]{Example}
\newenvironment{exa}{\begin{example} \rm }{ \end{example}}
\newtheorem{remark}[thm]{Remark}

\newenvironment{rmk}{\begin{remark} \rm }{\hfill $\Box$ \end{remark}}
\newenvironment{prf}{\noindent {\it Proof:} \ }{\hfill $\Box$}

\newcommand\od{\mathrm{d}}
\newcommand\p{\partial}

\newcommand{\nn}{\nonumber}

\newcommand{\al}{\alpha}

\newcommand{\dt}{\delta}

\newcommand{\res}{\mathrm{Res}}

\newcommand\Z{\mathbb{Z}}
\newcommand\C{\mathbb{C}}

 \newcommand\cB{\mathcal{B}}

\newcommand\cE{\mathcal{E}}

\newcommand{\set}[1]{\left\{#1\right\}}

\newcommand\ra{\right\rangle}
\newcommand\la{\left\langle}

\newcommand{\bm}[1]{\mathbf{#1}}
\newcommand{\bt}{\bm{t}}  
\newcommand{\bs}{\bm{s}}

\allowdisplaybreaks \numberwithin{equation}{section}

\allowdisplaybreaks
\begin{document}

\title[Reduction properties of the KP-mKP hierarchy]{Reduction properties of the KP-mKP hierarchy}
\author{Lumin Geng, Jianxun Hu, Chao-Zhong Wu}
\dedicatory {School of Mathematics, Sun Yat-sen University, Guangzhou 510275, P. R. China \\
Email address: genglm@mail2.sysu.edu.cn, stsjxhu@sysu.edu.cn,  wuchaozhong@sysu.edu.cn}

\begin{abstract}
The so-called KP-mKP hierarchy, which was introduced recently via pseudo-differential operators with two derivations,  can be reduced to the Kadomtsev-Petviashvili (KP), the modified KP (mKP) and the two-component BKP hierarchies. In this note, we continue to study reductions properties of the KP-mKP hierarchy, including its $(n,m)$-reduction and its reduction to a certain extended $r$-reduced KP hierarchy (the $r$-th Gelfand-Dickey together with its wave function). As a byproduct, we show that the Hirota equations of the extended $r$-reduced KP hierarchy follow from those of the mKP hierarchy, which confirms a conjecture of Alexandrov on the open KdV hierarchy in [{\it J. High Energy Phys. 2015}].
\end{abstract}
\maketitle

\section{Introduction}

Integrable hierarchies of Kadomtsev-Petviashvili (KP) type (see \cite{DJKM-KPtype,DKJM-KPBKP,DKJM-redKP,JM} for example) and their reduction properties have attracted great interest in the area of mathematical physics. Such kind of integrable hierarchies usually have Lax formalisms via pseudo-differential operators with a single derivation, and can be expressed as bilinear equations of Baker-Akhiezer functions or of tau functions. Among them, the two-component BKP (2-BKP) hierarchy \cite{DJKM-KPtype,JM} was later characterized in a fairly symmetric form by Shiota  \cite{ST1988} via certain pseudo-differential operators with two derivations (see also \cite{GHW2023}). With the help of pseudo-differential operators involving two derivations, a so-called KP-mKP hierarchy was introduced by the authors recently \cite{GHW2024}. This hierarchy has two tau functions, and it is represented into either of two different bilinear equations of its (adjoint) Baker-Akhiezer functions. Moreover, the KP-mKP hierarchy can be considered as a subhierarchy of the dispersive Whitham hierarchy \cite{KIM1994,SB2008,WZ2016} associated with the Riemann sphere with $\infty$ and one movable point marked, and it admits a B\"{a}cklund transformation such that $\infty$ and the movable point are changed to another movable point and $\infty$ respectively.

Starting from two pseudo-differential operators with derivations $\p_1$ and $\p_2$ of the form
\[
	\Phi_1=1+\sum_{i\ge1}a_{1,i} \p_{1}^{-i},\quad
	\Phi_2=e^{\beta }\Big(1+\sum_{i\ge1}a_{2,i}\p_{2}^{-i}\Big),
\]
the KP-mKP hierarchy is defined by the following evolutionary equations:
\begin{align}
\frac{\p \Phi_1}{\p t_{1, k}}=-\left(\Phi_1\p_1^k\Phi_1^{-1}\right)_{<0}\Phi_1,&\quad
\frac{\p \Phi_2}{\p t_{2, k}}=-\left(\Phi_2\p_2^k\Phi_2^{-1}\right)_{<1}\Phi_2, \label{re-Phinut10} \\
\frac{\p \Phi_1}{\p t_{2, k}}=\left(\Phi_2\p_2^k\Phi_2^{-1}\right)_{\geq1}\la\Phi_1\ra,&\quad
\frac{\p \Phi_2}{\p t_{1, k}}=\left(\Phi_1\p_1^k\Phi_1^{-1}\right)_{\geq0}\la\Phi_2\ra, \label{re-Phinut20} \\
e^\beta\p_1 e^{-\beta}\Phi_1\p_2\Phi_1^{-1}&=\p_2\Phi_2\p_1\Phi_{2}^{-1} \label{re-Phinumu0}
\end{align}
with  $k=1,2,3,\cdots$. Here the subscripts ``$< p$'' and ``$\ge p$'' mean truncations of  pseudo-differential operators, and the symbol ``$B\la A\ra$'' means that the differential operator $B$ acting on the coefficients of the operator $A$ (see Section~2 below for details).
It is known that the KP-mKP hierarchy can be reduced to the KP, the modified KP (mKP) and the 2-BKP hierarchies. In this note, let us continue to study reductions properties of the KP-mKP hierarchy.

Firstly, given two arbitrary positive integers $n$ and $m$, we assume the KP-mKP hierarchy \eqref{re-Phinut10}--\eqref{re-Phinumu0} to satisfy
\begin{align}\label{re-W-0ret}
	\left(\frac{\p}{\p t_{1,n}}+\frac{\p}{\p t_{2,m }}\right)\Phi_\nu=0,\quad \nu\in\{1,2\}.
\end{align}
Under this constraints, we will show that the KP-mKP hierarchy \eqref{re-Phinut10}--\eqref{re-Phinumu0}
is reduced to the following bilinear equations of (adjoint) Baker-Akhiezer functions:
\begin{equation}
	\res_z\left( z^{n l-1}w_1(\bt_1,\bt_2; z)w_1^{\dag}(\bt_1',\bt_2'; z) \right)=\res_z \left(z^{m l-1}w_2(\bt_1,\bt_2; z)w_2^{\dag}(\bt_1',\bt_2'; z)\right),\quad l\ge0,\label{nmred0}
\end{equation}
with $(\bt_1,\bt_2)$ and  $(\bt_1',\bt_2')$ being arbitrary time variables. Here $\bt_\nu=(t_{\nu,1},t_{\nu,2},t_{\nu,3},\dots)$. As to be seen, the case $m=1$ coincides with the constrained KP (cKP$_{n,1}$) hierarchy \cite{Cheng,DLA1995}; when $n$ is even and $m=2$, this $(n,m)$-reduced hierarchy \eqref{nmred0} can be further reduced to the Drinfeld-Sokolov hierarchies of type D \cite{DS,LWZ2011}.

Our another motivation is to study the open KdV hierarchy and its generalizations in the context  of the KP-mKP hierarchy. It is shown by Buryak \cite{Bur2015,Bur2016} that a special solution of  the Burgers-KdV hierarchy (an extension of the open KdV hierarchy with descendants) gives the generating function of intersection numbers on the moduli space of Riemann surfaces with boundary, constructed by Pandharipande, Solomon and Tessler \cite{PST}. Recently, an open $r$-spin theory was developed by Buryak, Clader and Tessler \cite{BCT2019,BCT2022,BCT2023,BCT2024}, and the open intersection numbers are governed by a generalization of the Burgers-KdV hierarchy, say, the so-called extended $r$-reduced KP hierarchy proposed in \cite{BY}.
More exactly, the $r$-reduced KP hierarchy is defined by
\begin{equation}\label{oGD0}
\frac{\p L}{\p s_{k}}=\left[(L^{k/r})_{\ge0},L\right], \quad
 \frac{\p \psi}{\p s_{k}}=(L^{k/r})_{\geq0}(\psi), \qquad k\in\Z_{\ge1},
\end{equation}
where $L$ is a differential operator of the form
\[
L=\p^r+u_1 \p^{r-2}+\dots+u_{r-2}\p+u_{r-1}, \quad \p=\frac{\mathrm{d}}{\mathrm{d}x},
\]
and $\psi$ is an unknown function.
Tau functions of the open KdV and its generalizations were studied from various points of view; see for instance \cite{Ale2015,Ale2015b,AB,BY,Ke,YZ2021}. In particular, Alexandrov \cite{Ale2015} conjectured that all Hirota equations satisfied by the tau functions of the open KdV hierarchy (and of the Burgers-KdV hierarchy \cite{Bur2015,Bur2016}) follow from the bilinear equation of the mKP hierarchy.

Given an integer $r\ge2$, let us assign to the KP-mKP hierarchy \eqref{re-Phinut10}--\eqref{re-Phinumu0} the following constraints:
\begin{equation}\label{redGD0}
(\Phi_1\p_1^r\Phi_1^{-1})_{<0}=0, \quad \Phi_2=e^\beta, \quad \p_2(\beta)=0.
\end{equation}
As to be seen, under these constraints the KP-mKP hierarchy is reduced to the extended $r$-reduced KP hierarchy \eqref{oGD0} (the Burgers-KdV hierarchy whenever $r=2$) with $\psi=e^\beta$ and the time variables $s_k=t_{1,k}$ for $k\ge1$. Accordingly, the two tau functions $\tau_1$ and $\tau_2$ of  the KP-mKP hierarchy are reduced to that of the  extended $r$-reduced KP hierarchy, of which the bilinear equations will be derived. As a byproduct, it will be shown that the  extended $r$-reduced KP hierarchy is indeed a reduction of the mKP hierarchy, and that the Hirota equations of its tau functions $\tau_1$ and $\tau_2$ follow from the bilinear equation of the latter. In particular, by taking $r=2$, one sees that Alexandrov's conjecture \cite{Ale2015} on the open KdV hierarchy is valid. Our method is based on the equivalent formalisms of the mKP hierarchy \cite{JM,KvdL} and the fact that the first tau function $\tau_1$ of the  extended $r$-reduced KP hierarchy is independent of the time variables $s_{r l}$ with $l\ge1$ (see Section~4 below).

This note is organized as follows. In Section~2, we recall some notations of pseudo-differential operators containing two derivations, as well as the definition of the KP-mKP hierarchy. In Section~3, we consider the $(n,m)$-reduction of the KP-mKP hierarchy, constrained by \eqref{re-W-0ret}.  In Section~4, we study the reduction of the KP-mKP hierarchy to the  extended $r$-reduced KP hierarchy, and clarify the relationship between the latter and the mKP hierarchy. The final section is dedicated to some remarks on subhierarchies of the dispersive Whitham hierarchy associated with the Riemann sphere with two movable points marked.

\section{Pseudo-differential operators and the KP-mKP hierarchy}
In this section, we review the notations of pseudo-differential operators involving two derivations, as well as the definition of the KP-mKP hierarchy introduced in \cite{GHW2024}.

\subsection{Pseudo-differential operators with two derivations}
Let $\cB$ be a commutative associative algebra of complex functions depending smoothly on two variables $x$ and $y$. The derivations $\p_1=\od/\od x$ and $\p_2=\od/\od y$ acting on $\cB$ commute with each other. We consider the following algebra of pseudo-differential operators:
\begin{equation*}\label{}
	\cE=\left\{ \sum_{i\le m}\sum_{j\le n}f_{i j} \p_1^i \p_2^j\mid f_{i j}\in\cB; \, m,n\in\Z \right\},
\end{equation*}
whose product is defined by
\begin{equation*}\label{}
	f \p_1^i\p_2^j\cdot g \p_1^p \p_2^q=\sum_{r,s\geq0}\binom{i}{r}\binom{j}{s} f\, \p_1^r\p_2^s(g)\cdot
	\p_1^{i+p-r}\p_2^{j+q-s}
\end{equation*}
with $f,\,g\in\cB$ and
\[
\binom{i}{r}=\frac{i(i-1)\dots(i-r+1)}{r!}.
\]
In $\cE$ there is a Lie bracket defined by the commutator, say,
\[
[A, B]=A B-B A, \quad A, B\in\cE.
\]

For $A=\sum_{i\le m}\sum_{j\le n}  {f}_{i j} \p_1^i\p_2^j \in\cE$, its residues mean
\begin{equation}\label{resA}
\res_{\p_1}A=\sum_{j\le n}  {f}_{-1, j} \p_2^j, \quad \res_{\p_2}A=\sum_{i\le m}  {f}_{i, -1} \p_1^i,
\end{equation}
and its adjoint operator is
\begin{equation}\label{star}
A^*=\sum_{i\le m}\sum_{j\le n} (-1)^{i+j}  \p_1^i\p_2^j f_{i j}.
\end{equation}
Note that ``$*$'' gives an anti-automorphism on $\mathcal{E}$, and that
\begin{equation}
(\res_{\p_\nu} A)^*=-\res_{\p_\nu} A^*, \quad \nu\in\set{1,2}.
\end{equation}
The action on $A$ by a given differential operator $D=\sum_{r,s\ge0}  {g}_{r s} \p_1^r\p_2^s\in\cE$ is denoted by
\begin{equation}\label{exDA}
D\la{A}\ra=\sum_{i,j} \left(\sum_{r,s\ge0}  {g}_{r s} \p_1^r \p_2^s (f_{i j})\right) \p_1^i\p_2^j,
\end{equation}
which means that the differential operator $D$ acts on each coefficient of $A$.

In particular, whenever the operator $A\in\cE$ contains powers in only $\p_1$ or $\p_2$, then the above notations agree with those for pseudo-differential operators with a single derivation. In this case, for $A=\sum_{i}f_i\p_\nu^i\in\mathcal{E}$ and $m\in\Z$, one has the following truncations:
\begin{equation}\label{cutop}
A_{\ge m}=\sum_{i\ge m} f_i \p_\nu^i, \quad A_{< m}=\sum_{i<m} f_i \p_\nu^i.
\end{equation}
The following result is useful below.
\begin{lem}[see, for example, \cite{DKJM-KPBKP}] \label{thm-exres}
	For any pseudo-differential operators $F, G\in\mathcal{E}$ that contain only powers in $\p_\nu$ with $\nu\in\{1,2\}$, the following equality holds true:
	\begin{equation*}
		\res_{z}\left(F e^{z x_\nu }\cdot G^* e^{-z x_\nu}\right)=\res_{\p_\nu} (F G).
	\end{equation*}
	Here $x_1=x$, $x_2=y$, $\p_\nu^i  e^{z x_\nu }=z^i  e^{z x_\nu }$ and $\res_{z}\left(\sum_i f_i z^i\right)=f_{-1}$.
\end{lem}

\subsection{The KP-mKP hierarchy}

Consider two pseudo-differential operators of the form:
\begin{align}
	\Phi_1=1+\sum_{i\ge1}a_{1,i} \p_{1}^{-i},\quad
	\Phi_2=e^{\beta }\Big(1+\sum_{i\ge1}a_{2,i}\p_{2}^{-i}\Big), \label{re-Phi120}
\end{align}
where $a_{\nu,i},\beta\in\cE$. These two operators are invertible, then one has the following  pseudo-differential operators
\begin{equation}
	P_1=\Phi_1\p_1\Phi_1^{-1}=\p_1+\sum_{i\ge1}u_{1,i}\p_{1}^{-i},\quad
	P_2=\Phi_2\p_2\Phi_2^{-1}=\p_2-\p_2(\beta)+\sum_{i\ge1}u_{2,i}\p_{2}^{-i}. \label{exP2}
\end{equation}
Note that the coefficients $u_{\nu,i}$ are differential polynomials belonging to the ring
\[
\mathcal{R}_\nu=\C\left[ \dt_{\nu 2}\p_\nu^m(\beta),\, \p_\nu^m(a_{\nu,k}) \mid m\ge0;\, k\ge1 \right].
\]
Here  $\dt$ stands for the Kronecker delta symbol. In particular, one has
\begin{equation}\label{u12}
u_{\nu,1}=-\p_\nu\left(a_{\nu,1}\right), \quad \nu\in\{1,2\}.
\end{equation}

Suppose that the operators $\Phi_\nu$ satisfy the condition
\begin{equation}
e^\beta\p_1 e^{-\beta}\Phi_1\p_2\Phi_1^{-1}=\p_2\Phi_2\p_1\Phi_{2}^{-1}, \label{Phinumu}
\end{equation}
then the KP-mKP hierarchy is composed of the following evolutionary equations:
\begin{align}
\frac{\p \Phi_1}{\p t_{1, k}}=-(P_1^k)_{<0}\Phi_1,&\quad
\frac{\p \Phi_2}{\p t_{2, k}}=-(P_2^k)_{<1}\Phi_2, \label{Phinut1} \\
\frac{\p \Phi_1}{\p t_{2, k}}=(P_2^k)_{\geq1}\la\Phi_1\ra,& \quad
 \frac{\p \Phi_2}{\p t_{1, k}}=(P_1^k)_{\geq0}\la\Phi_2\ra, \label{Phinut2}
\end{align}
where $k\in\Z_{\geq1}$. It is known that these equations are compatible, and that they can be reduced to the KP hierarchy and the modified KP hierarchy; see \cite{GHW2024} for details. Observe that ${\p}/{\p t_{\nu,1}}=\p_{\nu}$, and in what follows, we will just take
\[
t_{1,1}=x, \quad t_{2,1}=y.
\]

Let $H$ denote either side of \eqref{Phinumu}, then it is a differential operator of the form
\begin{align}\label{re-exH}
	H=\p_1\p_2-\p_1(\beta)\p_2-\rho,
\end{align}
where
\begin{align}
\rho=\p_2(a_{1,1})=\p_1\p_2(\beta)+\p_1(a_{2,1}).\label{re-rho}
\end{align}
It can be verified that (see Lemma~3.5 of \cite{GHW2024})
\begin{equation}\label{re-exH120}
\p_2\la\Phi_1\ra=e^{\beta}\p_1^{-1}e^{-\beta}\rho \Phi_1, \quad
\p_1\la\Phi_2\ra=\p_2^{-1}\left(\rho+\p_1(\beta)\p_2\right) \Phi_2.
\end{equation}

Given a solution of the KP-mKP hierarchy \eqref{Phinumu}--\eqref{Phinut2}, its Baker-Akhiezer functions read
\begin{equation}\label{wavef}
	w_\nu=w_\nu(\bt_1, \bt_2; z)=\Phi_\nu e^{\xi(\bt_\nu;z)}, \quad \nu\in\{1,2\},
\end{equation}
where $\bt_\nu=(t_{\nu,1},t_{\nu,2},t_{\nu,3},\dots)$ and
\[
\xi(\bt_\nu; z)=\sum_{k\in\Z_{\geq1}}
t_{\nu,k} z^k.
\]
These functions satisfy
\begin{equation}\label{wt}
\frac{\p w_\nu}{\p t_{\mu,k}}=(P_\mu^k)_{\ge \mu-1}w_\nu, \quad \mu,\nu\in\{1,2\}, ~ k\in\Z_{\ge1}.
\end{equation}
On the other hand, the adjoint Baker-Akhiezer functions of the KP-mKP hierarchy are defined by:
\begin{equation}\label{adwavef}
w_1^\dag(\bt_1,\bt_2;z) =\left(\p_1\Phi_1^{-1}e^{\beta}\p_1^{-1}e^{-\beta}\right)^*e^{-\xi(\bt_1;z)}, \quad
w_2^\dag(\bt_1,\bt_2;z) =\left(\p_2\Phi_2^{-1}\p_2^{-1}\right)^*e^{-\xi(\bt_2;z)}.
\end{equation}
It is easy to see the following equalities (recall \eqref{re-exH}):
\begin{equation}\label{}
H w_\nu=0, \quad H^* w_\nu^\dag=0, \qquad \nu\in\{1,2\}.
\end{equation}

\begin{thm}[\cite{GHW2024}]\label{thm-ble2}
(I) The KP-mKP hierarchy \eqref{Phinumu}--\eqref{Phinut2} is equivalent to the following bilinear equation
\begin{equation}\label{exble3}
\res_z\left( z^{-1} w_1(\bt_1, \bt_2; z)w_1^{\dag}(\bt_1', \bt_2'; z) \right)=\res_z \left(z^{-1}w_2(\bt_1, \bt_2; z)w_2^{\dag}(\bt_1', \bt_2'; z)\right)
\end{equation}
with arbitrary time variables $(\bt_1, \bt_2)$ and $(\bt_1', \bt_2')$.\\
(II) There exist two tau functions $\tau_1$ and $\tau_2=e^\beta\tau_1$ of $(\bt_1,\bt_2)$ such that
  \begin{align}
	w_1(\bt_1,\bt_2;z)=&\frac{\tau_1(\bt_1-[z^{-1}],\bt_2)}{\tau_1(\bt_1,\bt_2)} e^{\xi(\bt_1;z)},\quad w_1^{\dag}(\bt_1,\bt_2;z)=\frac{\tau_2(\bt_1+[z^{-1}],\bt_2)}{\tau_2(\bt_1,\bt_2)} e^{-\xi(\bt_1;z)},\label{w1tau} \\ w_2(\bt_1,\bt_2;z)=&\frac{\tau_2(\bt_1,\bt_2-[z^{-1}])}{\tau_1(\bt_1,\bt_2)} e^{\xi(\bt_2;z)},\quad w_2^{\dag}(\bt_1,\bt_2;z)=\frac{\tau_1(\bt_1,\bt_2+[z^{-1}])}{\tau_2(\bt_1,\bt_2)}e^{-\xi(\bt_2;z)}, \label{w2tau}
\end{align}
where $[z^{-1}]=\left(\frac{1}{z}, \frac{1}{2z^2}, \frac{1}{3z^3}, \dots\right)$.
\end{thm}

\section{The $(n,m)$-reduction of the KP-mKP  hierarchy }

In this section, given two positive integers $n$ and $m$, we assign to the KP-mKP hierarchy the following constraints
\begin{align}\label{nmred}
	\left(\frac{\p}{\p t_{1,n}}+\frac{\p}{\p t_{2,m }}\right)\Phi_\nu=0,\quad \nu\in\{1,2\},
\end{align}
and study the reduced evolutionary equations.

\subsection{Lax equations}
By virtue of \eqref{Phinut1}--\eqref{Phinut2}, the constraints \eqref{nmred} are equivalent to
\begin{align}\label{nmredPhi}
(P_1^{n})_{<0}=(P_2^{m})_{\geq1}\la\Phi_1\ra\Phi_1^{-1},\quad(P_2^{m})_{<1}=(P_1^{n})_{\geq0}\la\Phi_2\ra\Phi_2^{-1}.
\end{align}
\begin{defn}
The equations \eqref{Phinumu}--\eqref{Phinut2} constrained by the conditions \eqref{nmredPhi} compose the $(n,m)$-reduced KP-mKP hierarchy.
\end{defn}

Let us give more details of this reduced hierarchy. Denote
\begin{equation}\label{Lnm}
L=(P_1^{n})_{\geq0}+(P_2^{m})_{\geq1},
\end{equation}
which takes the form
\[
L=\p_1^n+\sum_{i=1}^{n-1} u_i\p_1^{n-i-1}+\p_2^m+\sum_{j=1}^{m-1}v_j\p_2^{m-j}.
\]
In particular, one has (cf. \eqref{u12})
\[
u_1=-n\p_1(a_{1,1}), \quad v_1=-m\p_2(\beta).
\]
The $(n,m)$-reduction yields the following Lax equations via pseudo-differential operators of two derivations:
\begin{align}\label{nmLt1}
\frac{\p L}{\p t_{1,k}}=&\left[(P_1^k)_{\ge0}, P_1^n\right]_{\ge0}+\left[(P_1^k)_{\ge0}\la \Phi_2\ra\Phi_2^{-1}, P_2^m\right]_{\ge1}, \\
\frac{\p L}{\p t_{2,k}}=&\left[(P_2^k)_{\ge1}\la \Phi_1\ra\Phi_1^{-1}, P_1^n\right]_{\ge0}+\left[(P_2^k)_{\ge1}, P_2^m\right]_{\ge1} \label{nmLt2}
\end{align}
with $k\in\Z_{\ge1}$. Note that these equations may not give all evolutionary equations of the whole $(n,m)$-reduced KP-mKP hierarchy, as to be seen in the following example.

\begin{exa}
When $m=1$, let
\[
L_1=P_1^n.
\]
One has, due to \eqref{nmredPhi} and \eqref{re-exH120},
\[
L_1=\p_1^n+\sum_{i=1}^{n-1} u_i\p_1^{n-i-1}+\left(\p_1-\p_1(\beta)\right)^{-1}\rho.
\]
Hence the
$(n,1)$-reduced KP-mKP hierarchy reads (recall $\p/\p t_{2,k}=-\p/\p t_{1,n k}$)
\begin{equation}\label{}
\frac{\p L_1}{\p t_{1,k}}=\left[ (L_1^{k/n})_{\ge0}, L_1\right], \quad k\in\Z_{\ge1}.
\end{equation}
This is just the $(n,1)$-constrained KP (cKP$_{n,1}$) hierarchy, and it gives the nonlinear Shr\"{o}dinger hierarchy and the Yajima--Oikawa hierarchy whenever $n=1$ and $2$ respectively; see \cite{Cheng,WZ2016,YO} and references therein. However, for the case $n=m=1$, the operator given in \eqref{Lnm} reads $L=\p_1+\p_2$, and the equations \eqref{nmLt1}--\eqref{nmLt2} are trivial.
\end{exa}

\begin{exa}
When $n=m=2$, let
\[
L_\nu=P_\nu^2, \quad \nu\in\{1,2\}.
\]
By using \eqref{nmredPhi} and \eqref{re-exH120}, it is easy to see
\begin{align}
L_1=&\p_1^2-2(a_{1,1})_x-\beta_y e^{\beta}\p_1^{-1}e^{-\beta}\rho +e^{\beta}\p_1^{-1}e^{-\beta}(\rho_y-\beta_y\rho)+e^{\beta}\p_1^{-1}\rho\p_1^{-1}e^{-\beta}\rho, \label{re-2P1}\\
L_2=&\p_2^2-2\beta_y\p_2-2(a_{1,1})_x + \beta_{x x}+\p_2^{-1}(\rho_x-\beta_{x x y})+\left(\beta_{x }+\p_2^{-1}(\rho-\beta_{x y}) \right)^2.  \label{re-2P2}
\end{align}
Here the subscripts $x$ and $y$ stand for the partial derivatives with respect to them, and $\rho=(a_{1,1})_y$ given in \eqref{re-rho}.
Denote $\al=a_{1,1}$, then we have
\[
L=(L_1)_{\ge0}+(L_2)_{\ge1}=\p_1^2-2\al_x +\p_2^2-2\beta_y \p_2.
\]
By using \eqref{nmLt1}--\eqref{nmLt2}, one can write down some explicit equations as follows:
\begin{align*}
\frac{\p\al}{\p t_{1,2}}=&2 \al_y\beta_y-\al_{y y}, \\
\frac{\p\beta}{\p t_{1,2}}=&\beta_x^2+\beta_{x x }-2\al_{x}, \\
\frac{\p\al}{\p t_{1,3}}=& -\frac{3}{2}\left( \al_x^2+\al_y^2+\al_{x y}\beta_y +\al_{y y}\beta_x -2 \al_y \beta_x \beta_y -\frac{1}{6}\al_{x x x}-\frac{1}{2}\al_{x y y}\right), \\
\frac{\p\beta}{\p t_{1,3}}=& -\frac{3}{2}\al_{x x}+\frac{3}{2}\al_{y y}-3\al_x\beta_x -3\al_y\beta_y+\beta_{x x x}+3\beta_{ x}\beta_{x x} +\beta_x^3,\\
\frac{\p\al}{\p t_{2,3}}=&  -\frac{3}{2}\left( 2 \al_x-\beta_{x x}-\beta_x^2 +\beta_{y y}-\beta_y^2 \right)\al_y -3\al_{y y}\beta_y +\al_{y y y}, \\
\frac{\p\beta}{\p t_{2,3}}=& -3\al_x\beta_y-3\al_y\beta_x+\frac{1}{4}\beta_{y y y}-\frac{1}{2}\beta_y^3+\frac{3}{4}\beta_{x x y}+\frac{3}{2}\beta_x^2\beta_y+\frac{3}{2}\beta_x\beta_{x y}+\frac{3}{2}\beta_{x x}\beta_y.
\end{align*}
\end{exa}

\subsection{Bilinear equations}
Let us represent the $(n,m)$-reduced KP-mKP  hierarchy in the form of bilinear equations.

\begin{lem}\label{thm-Lw}
For the $(n,m)$-reduced KP-mKP hierarchy, the differential operator $L$ given in \eqref{Lnm} and the  Baker-Akhiezer functions defined by \eqref{wavef} satisfy
\begin{equation}\label{}
  L w_1=z^n w_1, \quad L w_2=z^m w_2.
\end{equation}
\end{lem}
\begin{prf}
By using \eqref{wt} and \eqref{nmred}, we have
\begin{align*}\label{}
L w_1=&\left((P_1^{n})_{\geq0}+(P_2^{m})_{\geq1} \right)w_1 \\
=&\left(\frac{\p}{\p t_{1,n}}+ \frac{\p}{\p t_{2,m}} \right)\left(\Phi_1 e^{\xi(\bt_1;z)}\right) \\
=&\Phi_1 \left(\frac{\p}{\p t_{1,n}}+ \frac{\p}{\p t_{2,m}} \right)e^{\xi(\bt_1;z)} \\
=&z^n w_1.
\end{align*}
The case for $w_2$ is similar. The lemma is proved.
\end{prf}

\begin{thm}\label{thm-nmble}
The $(n,m)$-reduced KP-mKP hierarchy is equivalent to the following bilinear equations
\begin{align}
\res_z\left( z^{n l-1}w_1(\bt_1,\bt_2; z)w_1^{\dag}(\bt_1',\bt_2'; z) \right)&=\res_z \left(z^{m l-1}w_2(\bt_1,\bt_2; z)w_2^{\dag}(\bt_1',\bt_2'; z)\right), \quad l\in\Z_{\geq0}, \label{nmble}
\end{align}
with arbitrary time variables $(\bt_1,\bt_2)$ and $(\bt_1',\bt_2')$.
\end{thm}
\begin{prf}
For the $(n,m)$-reduced KP-mKP  hierarchy, let the differential operators $L^l$ act on the bilinear equation \eqref{exble3}, then we derive the bilinear equations \eqref{nmble} with the help of Lemma~\ref{thm-Lw}.

Conversely, since the case $l=0$ of \eqref{nmble} is just the bilinear equation of the KP-mKP hierarchy, then it suffices to deduce \eqref{nmredPhi} from \eqref{nmble}.

For $i\in\Z_{\ge0}$, let $\p_1^{i}$ act on both sides of \eqref{nmble} and take $(\bt_1',\bt_2')=(\bt_1,\bt_2)$, then by using \eqref{wavef}, \eqref{adwavef} and Lemma \ref{thm-exres} we have
\[ \res_{\p_1}\left(\p_1^{i}\Phi_1\p_1^{n l-1}\cdot\p_1 \Phi_1^{-1}e^{\beta}\p_1^{-1}e^{-\beta}\right)
=\res_{\p_2}\left(\p_1^{i}\la\Phi_2\ra\p_2^{m l-1}\cdot \p_2 \Phi_2^{-1}\p_2^{-1}\right).
\]
That is
\begin{equation}\label{blePhi}
\res_{\p_1} \left(\p_1^{i}P_1^{n l}e^{\beta}\p_1^{-1}e^{-\beta} \right)
=\res_{\p_2}\res_{\p_1}\left( \p_1^{i}
P_2^{m l}\Phi_2\p_1^{-1}\Phi_2^{-1} \p_2^{-1} \right).
\end{equation}
In particular, the case $i=0$ implies
\begin{align}\label{P1nlbeta}
(P_1^{n l})_{\geq0}(e^{\beta})e^{-\beta} =(P_2^{m l})_{\ge0}(1).
\end{align}
For general $i$, by using the condition \eqref{Phinumu} and the equality \eqref{blePhi} one has
\begin{align*}
0=&\res_{\p_1} \p_1^{i}\left(P_1^{n l}e^{\beta}\p_1^{-1}e^{-\beta}- \res_{\p_2}
P_2^{m l}\Phi_1\p_2^{-1}\Phi_1^{-1} e^{\beta}\p_1^{-1}e^{-\beta} \right)
\\
=&\res_{\p_1} \p_1^{i}\left(P_1^{n l}e^{\beta}\p_1^{-1}e^{-\beta}-
(P_2^{m l})_{\ge0}\la\Phi_1\ra \Phi_1^{-1}  e^{\beta}\p_1^{-1}e^{-\beta} \right).
\end{align*}
Since $i$ runs over $\Z_{\ge0}$, then we obtain
\[
\left(P_1^{n l}e^{\beta}\p_1^{-1}e^{-\beta}-
(P_2^{m l})_{\ge0}\la\Phi_1\ra \Phi_1^{-1}  e^{\beta}\p_1^{-1}e^{-\beta} \right)_{<0}=0,
\]
namely,
\[
(P_1^{n l})_{<0}e^{\beta}\p_1^{-1}e^{-\beta}+(P_1^{n l})_{\ge0}(e^{\beta})\p_1^{-1}e^{-\beta}- \left((P_2^{m l})_{\ge1}\la\Phi_1\ra \Phi_1^{-1} + (P_2^{m l})_{\ge0}(1) \right) e^{\beta}\p_1^{-1}e^{-\beta}=0.
\]
This equality together with \eqref{P1nlbeta} leads to
\[
(P_1^{n l})_{<0}=(P_2^{m l})_{\geq1}\la\Phi_1\ra\Phi_1^{-1}.
\]
Take $l=1$, then the first part of \eqref{nmredPhi} is verified. The second part of \eqref{nmredPhi} can be verified in the same way. Therefore the theorem is proved.
\end{prf}

From the proof we also obtain the following result (cf.\eqref{nmred}).
\begin{cor}
For the $(n,m)$-reduced KP-mKP hierarchy, it holds that
\begin{equation}\label{nmred2}
	\left(\frac{\p}{\p t_{1,n l}}+\frac{\p}{\p t_{2,m l}}\right)\Phi_\nu=0,\quad \nu\in\{1,2\},~ l\in\Z_{\ge1}.
\end{equation}
\end{cor}

Taking Theorem~\ref{thm-nmble} and the second item of Theorem~\ref{thm-ble2} together, we arrive at the following result.
\begin{cor}
The $(n,m)$-reduced KP-mKP hierarchy can be represented as the following bilinear equations of tau functions:
\begin{align}
&\res_z\left( z^{n l-1}\tau_1(\bt_1-[z^{-1}],\bt_2; z)\tau_2(\bt_1'+[z^{-1}],\bt_2'; z)e^{\xi(\bt_1-\bt_1';z)} \right) \nn\\
=&\res_z \left(z^{m l-1}\tau_2(\bt_1,\bt_2-[z^{-1}]; z)\tau_1(\bt_1',\bt_2'+[z^{-1}]; z)e^{\xi(\bt_2-\bt_2';z)} \right), \quad l\in\Z_{\geq0}, \label{nmtauble}
\end{align}
with  arbitrary time variables $(\bt_1,\bt_2)$ and $(\bt_1',\bt_2')$.
\end{cor}

\begin{rmk}
In the KP-mKP hierarchy \eqref{Phinumu}--\eqref{Phinut2}, suppose that $\beta=0$ and
\[
\Phi_\nu^*=\p_\nu\Phi_\nu^{-1}\p_\nu^{-1}, \quad \nu\in\{1,2\},
\]
then the flows $\p/\p t_{\nu,k}$ with odd indices $k$ are well defined, and they compose the two-component BKP hierarchy \cite{DJKM-KPtype,GHW2023,ST1988}. In this case the (adjoint) Baker-Akhiezer functions (recall \eqref{wavef} and \eqref{adwavef}) satisfy
\[
w_\nu^\dag\left(\bt^{\mathrm{odd}};z\right) =\left(\p_\nu\Phi_\nu^{-1}\p_\nu^{-1}\right)^*e^{-\sum_{j\ge0}t_{\nu,2 j+1}z^{2 j+1}} =w_\nu \left(\bt^{\mathrm{odd}};-z\right),
\]
where $\bt^{\mathrm{odd}}=\left(\bt_1^{\mathrm{odd}},\bt_2^{\mathrm{odd}} \right)$ with $\bt_\nu^{\mathrm{odd}}=(t_{\nu,1},t_{\nu,3},t_{\nu,5},\dots)$. In the same way as above, for any positive integers $p$ and $q$, one can derive the bilinear equations of the so-called $(2p,2q)$-reduction of the two-component BKP hierarchy:
\[
\res_z\left( z^{2p l-1}w_1(\bt^{\mathrm{odd}}; z)w_1({\bt'}^{\mathrm{odd}}; - z) \right) =\res_z \left(z^{2q l-1}w_2(\bt^{\mathrm{odd}}; z)w_2({\bt'}^{\mathrm{odd}}; - z)\right), \quad l\in\Z_{\geq0},
\]
with arbitrary time variables $\bt^{\mathrm{odd}}$ and ${\bt'}^{\mathrm{odd}}$. In particular, the $(2p,2)$-reduction is equivalent to the Drinfeld-Sokolov hierarchy of type $D_{p+1}^{(1)}$ \cite{DS,LWZ2011} (cf. \cite{DKJM-redKP}).
\end{rmk}


\section{Reduction to the extended $r$-reduced KP hierarchy }

Given an integer $r\ge2$, we assume that the KP-mKP hierarchy \eqref{Phinumu}--\eqref{Phinut2} satisfies the following conditions:
\begin{equation}\label{redGD}
(\Phi_1\p_1^r\Phi_1^{-1})_{<0}=0, \quad \Phi_2=e^\beta, \quad \p_2(\beta)=0.
\end{equation}
Clearly, one has $P_2=e^{\beta}\p_2 e^{-\beta}=\p_2$. Moreover, the condition \eqref{Phinumu} implies that $\p_2\la\Phi_1\ra=0$. Hence, under the conditions \eqref{redGD}, the KP-mKP hierarchy is reduced to
\begin{equation}\label{oGD1}
\frac{\p \Phi_1}{\p s_{k}}=-(P_1^k)_{<0}\Phi_1, \quad
 \frac{\p e^\beta}{\p s_{k}}=(P_1^k)_{\geq0}(e^\beta),
\end{equation}
where $k\in\Z_{\ge1}$, and we write $s_k=t_{1,k}$ to simplify the notations.

Denote $L=P_1^r$, then it takes the form
\[
L=\p_1^r+u_1 \p_1^{r-2}+\dots+u_{r-2}\p_1+u_{r-1}.
\]
Note $P_1=L^{1/r}$,
Thus the equations \eqref{oGD1} can be rewritten as
\begin{equation}\label{oGD2}
\frac{\p L}{\p s_{k}}=\left[(L^{k/r})_{\ge0},L\right], \quad
 \frac{\p e^\beta}{\p s_{k}}=(L^{k/r})_{\geq0}(e^\beta), \qquad k\in\Z_{\ge1}.
\end{equation}
This system of evolutionary equations with $r=2$ is called the Burgers-KdV hierarchy, whose special solution gives the generating series of the intersection numbers on the moduli space of Riemann surfaces with boundary \cite{Bur2015,Bur2016,PST}.
For general $r$, the system \eqref{oGD2} is called the extended $r$-reduced KP hierarchy \cite{BY}, which plays an important role in the open/extended $r$-spin theories \cite{BCT2019,BCT2022,BCT2023,BCT2024}.

Let us proceed to represent the hierarchy \eqref{oGD2} into bilinear equations in the context of the KP-mKP hierarchy.

Similar as in \eqref{wavef} and \eqref{adwavef}, the Baker-Akhiezer function and the adjoint  Baker-Akhiezer function of the extended $r$-reduced KP hierarchy are defined respectively by
\begin{equation}\label{wavefoGD}
w(\bs; z)=\Phi_1 e^{\xi(\bs;z)}, \quad w^\dag(\bs; z)=\left(\p_1\Phi_1^{-1}e^\beta \p_1^{-1} e^{-\beta}\right)^* e^{-\xi(\bs;z)},
\end{equation}
where $\bs=(s_1,s_2,s_3,\dots)$.
\begin{thm}\label{thm-oGD}
(I) The (adjoint)  Baker-Akhiezer functions of the extended $r$-reduced KP hierarchy satisfy the following bilinear equations:
\begin{equation}\label{oGDble}
\res_z \left( z^{ r l-1}w(\bs;z)w^\dag(\bs';z) \right)
=\left\{\begin{array}{cl}
e^{\beta(\bs)-\beta(\bs')}, & \quad l=0; \\
\dfrac{\p e^{\beta(\bs)} }{\p s_{r l}}e^{-\beta(\bs')},& \quad l \ge1,
                                                       \end{array}
\right.
\end{equation}
with arbitrary time variables $\bs$ and $\bs'$.
\\
(II) Conversely, suppose that two functions of the form
\begin{equation}\label{}
w(\bs;z)=\left(1+\sum_{i\ge1}a_i(\bs)z^{-i}\right)e^{\xi(\bs;z)}, \quad w^\dag(\bs;z)=\left(1+\sum_{i\ge1}\tilde{a}_i(\bs)z^{-i}\right)e^{-\xi(\bs;z)}
\end{equation}
satisfy the bilinear equation \eqref{oGDble}, then they are the (adjoint)  Baker-Akhiezer functions of the extended $r$-reduced KP hierarchy.
\end{thm}
\begin{prf}
Thanks to the conditions \eqref{redGD}, the case $l=0$ of \eqref{oGDble} follows by taking $\Phi_2=e^{\beta}$ and $\bt_2=\bt_2'=0$ in the bilinear equation \eqref{exble3}, and the equalities \eqref{oGDble} with general $l$ follow by letting the differential operators $L^l=P_1^{r l}$ act on both sides of the case $l=0$.

Conversely, noting that $\p_1=\p/\p t_1=\p/\p s_1$, one introduces two pseudo-differential operators:
\[
\Phi=1+\sum_{i\ge1}a_i(\bs)\p_1^{-i}, \quad \tilde{\Phi}=1+\sum_{i\ge1}\p_1^{-i}\tilde{a}_i(\bs).
\]
Let $\p_1^i$ with $i\ge0$ act on both sides of \eqref{oGDble} and take $\bs'=\bs$, then according to Lemma~\ref{thm-exres} one has
\begin{equation}\label{eq1-oGD}
\res_{\p_1}\left( \p_1^i\Phi\p_1^{r l-1}\tilde{\Phi} \right)=
\left\{\begin{array}{cl}
\p_1^i\left( e^\beta \right)e^{-\beta},& l=0; \\
\p_1^i\left(\frac{\p e^\beta}{\p s_{r l}}\right)e^{-\beta},& \quad l\ge1.
                                                             \end{array}
                                                             \right.
\end{equation}
The case $l=0$ of \eqref{eq1-oGD} implies that
\[
\Phi\p_1^{-1}\tilde{\Phi}=\sum_{i\ge0}\p_1^{-i-1}\cdot \p_1^i\left(  e^\beta \right)e^{-\beta}
=e^{\beta}\p_1^{-1} e^{-\beta}.
\]
Hence we obtain
\begin{equation}\label{eq2-oGD}
\tilde{\Phi}=\p_1\Phi^{-1}e^{\beta}\p_1^{-1} e^{-\beta}.
\end{equation}
Thanks to this equality, for $i\ge0$ and $k\ge1$, now let $\p_1^i \frac{\p}{\p s_{k}}$ act on \eqref{oGDble} with $l=0$ and take $\bs'=\bs$, then by using Lemma~\ref{thm-exres} one has
\[
\res_{\p_1} \left( \p_1^i\left(\frac{\p \Phi}{\p s_{k}}+\Phi\p_1^k\right)\Phi^{-1}e^{\beta}\p_1^{-1}e^{-\beta} \right)= \p_1^i\left(\frac{\p e^\beta}{\p s_{k}}\right)e^{-\beta}.
\]
In the same way as above, we arrive at
\[
\left(  \left(\frac{\p \Phi}{\p s_{k}}+\Phi\p_1^k\right)\Phi^{-1}e^{\beta}\p_1^{-1}e^{-\beta} \right)_{<0}=\frac{\p e^\beta}{\p s_{k}}\p_1^{-1}e^{-\beta}.
\]
Then by considering the coefficients of $\p_1^{-1}$ and lower powers, we obtain
\begin{equation}\label{eq3-oGD}
\left(\Phi\p_1^{k}\Phi^{-1} \right)_{\ge0}(e^{\beta})=\frac{\p e^\beta}{\p s_{k} }, \quad
\left(  \left(\frac{\p \Phi}{\p s_k}+\Phi\p_1^k\right)\Phi^{-1} \right)_{<0}=0,
\end{equation}
of which the second equality implies
\[
\frac{\p \Phi}{\p s_k}=-\left(\Phi\p_1^k\Phi^{-1}\right)_{<0} \Phi.
\]

Finally, substituting \eqref{eq2-oGD} into \eqref{eq1-oGD} with $l\ge1$, and using \eqref{eq3-oGD}, we have
\[
\left(\p_1^i\Phi\p_1^{r l}\Phi^{-1} \right)_{\ge0}(e^{\beta})
= \p_1^i\left(\frac{\p e^\beta}{\p s_{r l}}\right)=\p_1^i \left(\Phi\p_1^{r l}\Phi^{-1} \right)_{\ge0}(e^{\beta}).
\]
Hence
\[
\left( \p_1^i\left(\Phi\p_1^{r l}\Phi^{-1}\right)_{<0}\right)_{\ge0}(e^\beta)=0, \quad i\ge0,
\]
and it follows that $\left(\Phi\p_1^{r l}\Phi^{-1}\right)_{<0}=0$.
Therefore the theorem is proved.
\end{prf}

By using the second item of Theorem~\ref{thm-ble2}, there are two tau functions $\tau_1(\bs)$ and $\tau_2(\bs)=\tau_1(\bs)e^{\beta(\bs)}$ such that the (adjoint) Baker-Akhiezer functions given in \eqref{wavefoGD} can be represented as
\begin{equation}\label{wavefoGD2}
w(\bs;z)=\frac{\tau_1(\bs-[z^{-1}] )}{\tau_1(\bs )} e^{\xi(\bs;z)},\quad w^\dag(\bs;z)=\frac{\tau_2(\bs+[z^{-1}] )}{\tau_2(\bs )} e^{-\xi(\bs;z)}.
\end{equation}

\begin{rmk} \label{rmk-tau1}
Clearly, the operator $\Phi$ solves the KP hierarchy, hence the function $\tau_1$ solves the bilinear equation of the KP hierarchy, namely,
\[
\res_z \left( \tau_1(\bs-[z^{-1}] )\tau_1(\bs'+[z^{-1}] )e^{\xi(\bs-\bs';z)}  \right)
=0.
\]
Moreover, by using \eqref{redGD} and \eqref{oGD1} one has
\[
\frac{\p\Phi}{\p s_{r l}}=-(\Phi\p_1^{r l}\Phi^{-1})_{<0}\Phi=0, \quad l\ge1.
\]
which implies that the tau function $\tau_1$ is independent of $\{s_{r l}\}_{l\ge1}$. As a consequence, the function $\tau_1$ is in fact the tau function of the $r$-th Gelfand-Dickey hierarchy \cite{DKJM-KPBKP,Dic}. In particular, when $r=2$ the function $\tau_1$ is the tau functions of the KdV hierarchy.
\end{rmk}

\begin{thm}\label{thm-tauoGD}
The extended $r$-reduced KP hierarchy is equivalent to the following bilinear equations of tau functions:
\begin{equation}\label{oGDbletau}
\res_z \left( z^{-1} \tau_1(\bs-[z^{-1}] )\tau_2(\bs'+[z^{-1}] )e^{\xi(\bs-\bs';z)}  \right)
=\tau_2(\bs )\tau_1(\bs' ),
\end{equation}
with arbitrary time variables $\bs$ and $\bs'$, and $\tau_1$ independent of $\{s_{r l}\}_{l\ge1}$.
\end{thm}
\begin{prf}
As explained before, the function $\tau_1$ is independent of $s_{r l}$ with $l\ge1$.
Substituting \eqref{wavefoGD2} into \eqref{oGDble}, we obtain the following result
\begin{equation}\label{oGDbletau2}
\res_z \left( z^{ r l-1}\tau_1(\bs-[z^{-1}] )\tau_2(\bs'+[z^{-1}] )e^{\xi(\bs-\bs';z)}  \right)
=\left\{\begin{array}{cl}
\tau_2(\bs )\tau_1(\bs' ), & \quad l=0; \\
\dfrac{\p \tau_2(\bs)}{\p s_{r l}}\tau_1(\bs' ) ,& \quad l \ge1.                                                  \end{array}
\right.
\end{equation}
Observe that the cases $l\ge1$ can be derived by taking the derivative of the case $l=0$ with respect to $s_{r l}$.
The theorem is proved.
\end{prf}

Note that the case $l=0$ of the bilinear equation \eqref{oGDbletau2} coincides with equation~(24) of \cite{KvdL}, which is one of the several equivalent formalisms of the (1st) mKP hierarchy (cf. \cite{JM}). Hence the following result is obtained.
\begin{cor} \label{thm-oGDmKP}
The extended $r$-reduced KP hierarchy \eqref{oGD2} is a reduction of the mKP hierarchy.
\end{cor}

For the case $r=2$, it is known that the tau functions $\tau_1=\tau_{WK}$ and $\tau_2=\tau_o$ of the open KdV hierarchy satisfy (see for instance \cite{Ale2015})
\begin{equation}\label{mKP}
\res_z \left( z\tau_2(\bs-[z^{-1}] )\tau_1(\bs'+[z^{-1}] )e^{\xi(\bs-\bs';z)}  \right)
=0,
\end{equation}
which is another formalism of the (1st) mKP hierarchy (see equation~(4) of \cite{KvdL}). In \cite{Ale2015} Alexandrov conjectured that all equations of the open KdV hierarchy (and of the Burgers-KdV hierarchy \cite{Bur2015,Bur2016})
follow from the bilinear equation of the mKP hierarchy. According to Theorem~\ref{thm-tauoGD} and Corollary~\ref{thm-oGDmKP}, the open KdV hierarchy is indeed a certain reduction of the mKP hierarchy. More exactly, any solution of the open KdV hierarchy fulfills not only the mKP hierarchy (the case $l=0$ of \eqref{oGDbletau2}) but also a series of bilinear equations (the cases $l\ge1$ of \eqref{oGDbletau2}); however, the equations with $l\ge1$ can be derived from the case $l=0$. Thus we arrive at the following result.
\begin{cor}
All bilinear equations of the tau functions $\tau_1$ and $\tau_2$ of the extended $r$-reduced KP hierarchy follow from those of the mKP hierarchy. In particular, the case $r=2$ means that Alexandrov's conjecture \cite{Ale2015} on the open KdV hierarchy is valid.
\end{cor}

\begin{exa}\label{re-exmkp}
Taking $r=2$ in \eqref{oGDbletau2}, one can derive the Hirota bilinear equations of the Burgers-KdV hierarchy. We recall the Hirota operators $D_{k}$ acting on a pair of functions $f(\bs)$ and $g(\bs)$ defined by
\begin{equation}\label{Dknu}
D_{k}f\cdot g= \left.\frac{\p }{\p \epsilon}\right|_{s=0} \left(\left.f\right|_{s_{k}\mapsto s_{k}+\epsilon} \left.g\right|_{s_{k}\mapsto s_{k}-\epsilon} \right).
\end{equation}
By a straightforward calculation, one can write down explicitly the following Hirota bilinear equations:
\begin{align}\label{re-hir000}
&\left({D_{1}}^2+D_2\right)\tau_1  \cdot\tau_2 =0,
\\
&
\left({D_{1}}^3-3D_1D_2-4D_3\right)\tau_1  \cdot\tau_2=0,\label{re-hir100}
\\
&\left(2D_4+D_2^2-D_1^2D_2\right)\tau_1 \cdot\tau_2=0,\label{re-hir2001}
\\
&
\left(D_1^4+8D_1D_3-6D_2^2+3D_1^2D_2\right)
\tau_1 \cdot\tau_2=0,
\label{re-hir0101}
\\
&\left(D_1^5+5D_1D_2^2-10D_1D_4-16D_5\right)
\tau_1 \cdot\tau_2=0, \label{re-hir1101}
\end{align}
These equations coincide with those of the mKP hierarchy listed in \cite{JM} (see also \cite{Ale2015}). Note that the equations \eqref{re-hir000} and \eqref{re-hir2001} lead to
\[
\left( D_2^2 + D_4 \right)
\tau_1 \cdot\tau_2=0,
\]
which implies (cf. the first equation in Section~5.2 of \cite{Bur2016})
\[
\frac{\p^2 \tau_2}{\p s_2^2}-\frac{\p \tau_2}{\p s_4}=0.
\]
\end{exa}

\begin{rmk}
For the case $r=2$, Yang and Zhou \cite{YZ2021} also introduced a tau function $\tau_{\mathrm{E}}$ of the Burgers-KdV hierarchy. This tau function satisfies
\[
\p_1\left(\log\frac{\tau_{\mathrm{E}}}{\tau_1}\right)=\p_1(\beta)
\]
with $\tau_1$ being the tau function of the KdV hierarchy (recall Remark~\ref{rmk-tau1}). One observes that, the tau function $\tau_{\mathrm{E}}$ agrees with $\tau_2=e^\beta \tau_1$ induced from that of the KP-mKP hierarchy, up to a freedom in the definition of tau functions.
\end{rmk}

\section{Concluding remarks}

Under certain conditions, the KP-mKP hierarchy defined via differential operators admits the $(n,m)$-reduction and the reduction to the extended $r$-reduced KP hierarchy, with integers $n,m\ge1$ and $r\ge2$. The bilinear equations of such reduced hierarchies have been obtained. In particular, it is shown that the equations of the extended $r$-reduced KP hierarchy can be obtained by those of the mKP hierarchy, which confirms the conjecture of Alexandrov on the open KdV hierarchy. We hope that these results would be helpful to understand integrable systems represented via pseudo-differential operators with more than one derivation, and to compute the intersection numbers on the moduli space of Riemann surfaces with boundary.

It is known that the KP-mKP hierarchy is a subhierarchy of the dispersive Whitham hierarchy 
associated to the Riemann sphere with the infinity $\infty$ and a movable point marked. Moreover, this hierarchy
admits a B\"{a}cklund transformation corresponding to a change of the marked points on the Riemann sphere: the infinity to a movable point, while the movable point to the infinity. Similarly, we can start from two pseudo-differential operators of the form
\[
\Phi_\nu=e^{\beta_\nu}\left(1+\sum_{i=1}^{\infty}a_{\nu,i} \p_{\nu}^{-i}\right)\in\cE, \quad \nu\in\{1,2\}, \label{WHm-Phi}
\]
where $a_{\nu,i}\in\cB$, and $\beta_\nu\in\cB\setminus\{0\}$ such that $\p_1\p_2(\beta_1-\beta_2)\ne0$.
Suppose that these operators satisfy
\[
e^{\beta_2}\p_1e^{-\beta_2}\Phi_1\p_2\Phi_1^{-1}=\p_1\p_2-\p_1(\beta_2)\p_2-\p_2(\beta_1)\p_1=e^{\beta_1}\p_2 e^{-\beta_1}\Phi_2\p_1\Phi_2^{-1},
\]
then the following evolutionary equations are well defined and compatible
\begin{align}
\frac{\p\Phi_{\mu}}{\p t_{\nu,k}}=&
\left\{ \begin{array}{cl}
-\left(\Phi_\nu\p_\nu^k\Phi_\nu^{-1}\right)_{<1}\Phi_\nu, &  \mu=\nu; \\
\left(\Phi_\nu\p_\nu^k\Phi_\nu^{-1}\right)_{\ge 1}\la\Phi_{\mu}\ra, & \mu\neq\nu,
\end{array}\right.
 \label{Phi12t}
\end{align}
where $\nu,\mu\in\{1,2\}$.
These equations in fact compose a subhierarchy of the dispersive Whitham hierarchy associated to the Riemann sphere with two distinct movable points marked. One can also represent this hierarchy to a bilinear equation of the form \eqref{exble3}. Moreover, via the transformation
\[
\left(\Phi_1, \Phi_2\right)\mapsto \left(e^{-\beta_1}\Phi_1, e^{-\beta_1}\Phi_2\right),
\]
the hierarchy \eqref{Phi12t} is converted to the KP-mKP hierarchy. Along this line, it is still unknown how to construct the dispersive Whitham hierarchy  with more than two points marked on the Riemann sphere. We will look for such a hierarchy and study its reduction properties elsewhere.

{\bf Acknowledgments.}
{\noindent \small The authors thank Chunhui Zhou for helpful discussions.
This work is partially supported by National Key R\&D Program of China 2023YFA10098001, NSFC No.\,12471243 and Guangzhou S\&T Program No. SL2023A04J01542. }



\begin{thebibliography}{99}




\bibitem{Ale2015} Alexandrov, A. Open intersection numbers, matrix models and MKP hierarchy. J. High Energy Phys. (2015), no. 3, 042, front matter+13pp.

\bibitem{Ale2015b} Alexandrov, A.
Open intersection numbers, Kontsevich-Penner model and cut-and-join operators.
J. High Energy Phys. 2015, no. 8, 028, front matter+24 pp.

\bibitem{AB} Aleshkin, K.; Belavin, V. Open minimal strings and open Gelfand-Dickey hierarchies.
J. High Energy Phys. 2019, no. 2, 043, front matter+21pp.


\bibitem{BY} Bertola, M.; Yang, D. The partition function of the extended r-reduced Kadomtsev-Petviashvili hierarchy. J. Phys. A 48 (2015), no. 19, 195205, 20pp.


\bibitem{Bur2015} Buryak, A.
Equivalence of the open KdV and the open Virasoro equations for the moduli space of Riemann surfaces with boundary.
Lett. Math. Phys. 105 (2015), no. 10, 1427--1448.

\bibitem{Bur2016}
Buryak, A.
Open intersection numbers and the wave function of the KdV hierarchy.
Mosc. Math. J. 16 (2016), no. 1, 27--44.

\bibitem{BCT2019} Buryak, A.; Clader, E.; Tessler, R. J. Closed extended $r$-spin theory and the Gelfand-Dickey wave function. J. Geom. Phys. 137 (2019), 132--153.

\bibitem{BCT2022} Buryak, A.; Clader, E.; Tessler, R. J. Open $r$-spin theory I: Foundations. Int. Math. Res. Not. IMRN 2022, no. 14, 10458--10532.

\bibitem{BCT2023} Buryak, A.; Clader, E.; Tessler, R. J.  Open $r$-spin theory III: A prediction for higher genus. J. Geom. Phys. 192 (2023), Paper No. 104960, 12 pp.


\bibitem{BCT2024} Buryak, A.; Clader, E.; Tessler, R. J. Open r-spin theory II: The analogue of Witten's conjecture for $r$-spin disks. J. Differential Geom. 128 (2024), no. 1, 1--75.


\bibitem{Cheng} Cheng, Y.
Constraints of the Kadomtsev-Petviashvili hierarchy.
J. Math. Phys. 33 (1992), no. 11, 3774--3782.


\bibitem{DJKM-KPtype} Date, E.; Jimbo, M.; Kashiwara, M.; Miwa, T. Transformation groups for soliton equations. IV. A new hierarchy of soliton equations of KP-type. Phys. D 4 (1981/82), no. 3, 343--365.

\bibitem{DKJM-KPBKP}
Date, E.; Kashiwara, M.; Jimbo, M.; Miwa, T. Transformation groups
for soliton equations. Nonlinear integrable systems---classical
theory and quantum theory (Kyoto, 1981), 39--119, World Sci.
Publishing, Singapore, 1983.

\bibitem{DKJM-redKP} Date, E.; Kashiwara, M.; Jimbo, M.; Miwa, T. Transformation groups for soliton equations--Euclidean Lie algebras and reduction of the KP hierarchy.
Publ. Res. Inst. Math. Sci. 18(1982), no. 3, 1077--1110.



\bibitem{DLA1995}Dickey, L. A.: On the constrained KP hierarchy. II. Lett. Math. Phys. 35 (1995), 229--236.

\bibitem{Dic}Dickey, L. A. Soliton Equations and Hamiltonian Systems. Singapore: World Scientific, 2003.


\bibitem{DS}Drinfel'd, V. G.; Sokolov, V. V.  Lie algebras and equations of Korteweg-de Vries type.  Current problems in mathematics 24, Itogi Nauki i
  Tekhniki, pages 81--180. Akad. Nauk SSSR, Vsesoyuz. Inst. Nauchn. i Tekhn.
  Inform., Moscow, 1984.



\bibitem{GHW2023}Geng, L.; Hu, J.; Wu, C.-Z. On Lax equations of the two-component BKP hierarchy. Phys. D 449 (2023), Paper No. 133748, 10 pp.

\bibitem{GHW2024}Geng, L.; Hu, J.; Wu, C.-Z. A KP-mKP hierarchy via pseudo-differential
operators with two derivations. Nonlinearity. 37 (2024), 095016, 34pp.


\bibitem{JM} Jimbo, M.; Miwa, T. Solitons and infinite-dimensional Lie algebras. Publ. Res. Inst. Math. Sci. 19 (1983), no.3, 943-1001.

\bibitem{KvdL} Kac, V. G.; van de Leur, J. W. Equivalence of formulations of the MKP hierarchy and its polynomial tau-functions. Jpn. J. Math. 13 (2018), no. 2, 235--271.

\bibitem{Ke} Ke, H.
On a geometric solution to open KdV and Virasoro. Adv. Math. (China)   46 (2017), no. 1, 91--96.

\bibitem{KIM1994}Krichever, I. M. The $\tau$-function of the universal Whitham hierarchy, matrix models and topological field theories. Comm. Pure Appl. Math. 47 (1994), no. 4, 437--475.



\bibitem{LWZ2011} Liu, S.-Q.; Wu, C.-Z.; Zhang, Y. On the Drinfeld-Sokolov hierarchies of~$D$ type.
Int. Math. Res. Not. IMRN (2011), no.8, 1952--1996.




\bibitem{PST}
Pandharipande, R.; Solomon, J. P.; Tessler, R. J.
Intersection theory on moduli of disks, open KdV and Virasoro.
Geom. Topol. 28 (2024), no. 6, 248--2567.

\bibitem{ST1988}Shiota, T. Prym Varieties and Soliton Equations.
Infinite-dimensional Lie algebras and groups (Luminy-Marseille, 1988), 407-448, Adv. Ser. Math. Phys., 7, World Sci. Publ., Teaneck, NJ, 1989.



\bibitem{SB2008} Szablikowski, B. M.; Blaszak, M. Dispersionful analog of the Whitham hierarchy. J. Math. Phys. 49 (2008), no.8, 082701, 20pp.











\bibitem{WZ2016}Wu, C.-Z.; Zhou, X. An extension of the Kadomtsev-Petviashvili hierarchy and its hamiltonian structures. J. Geom. Phys. 106 (2016), 327--341.

\bibitem{YZ2021}Yang, D.; Zhou, C.
On an extension of the generalized BGW tau-function.
Lett. Math. Phys. 111 (2021), no. 5, Paper No. 123, 23 pp.


\bibitem{YO}Yajima, N.; Oikawa, M.  Formation and interaction of Sonic-Langmuir solitons, Prog. Theor.
Phys., 56 (1976), No. 6, 1719--1739.

\end{thebibliography}
\end{document}